# Theoretical Investigation of Fermion Pairings in Three-band Extended Hubbard Model

Partha Goswami

*Department of Physics, D.B. College, University of Delhi, Kalkaji,Delhi-110019,India*

**Abstract.** The two-dimensional d-p model (or extended Hubbard model ) on a square lattice is investigated for fermion pairing by a slave boson method. The inter-site d-fermion interaction is introduced additionally. The momentum space counterpart of this interaction is taken to be separable and expanded in terms of basis functions corresponding to mixed symmetry states. The investigation leads to anomalous pairing of fermion fields only for pure d-wave symmetry. The charge and spin ordering gaps appear in the single- particle spectrum when d(x,y) state is taken into account.The non-Fermi liquid behavior is found to be the prevalent one generally. The study yields a good qualitative account of the nodal-antinodal dichotomy. Two gaps appear in the normal state spectrum; the larger one corresponds to the sum and the smaller to the difference of spin and charge ordering gaps. This two-gap scenario is in qualitative agreement with recent experimental finding.

## I. INTRODUCTION

The high temperature superconductivity in hole-doped cuprates is derived from doping the parent anti-ferromagnetic, charge-transfer insulators. It is now a common view[1-4] that the superconducting (SC) gap in such systems has robust d-wave symmetry.Therefore, in the SC state of these systems sharp, long-lived quasi-particle like excitations (QPE) remain possible near the nodal region centered around ($\pm\pi/2, \pm\pi/2$) where the gap is zero.In the anti-nodal sector centered around [($\pm\pi, 0$),($0,\pm\pi$)],on the other hand, QPEs are inconspicuous. This is signaled by the broadening of the QPE peak in the spectral function and decrease in their life-time.It was proposed by previous workers[2] that the antinodal spectral broadening is due to the coupling of electrons with the ($\pi,\pi$) magnetic excitations. In fact,the normal state properties of cuprates are highly anoma-lous, particularly, in the under-doped region. The nodal fermionic states possibly play significant role with regard to these properties. A great deal of efforts have been made on both theoretical[5-8] experimental[9-12] fronts to understand this role.

The conventional single-band Hubbard model, generally believed[13,14] to be adequate for the theoretical investigation of d-wave superconductivity in cuprates, has remained unsolved exactly till date except in one or infinite dimensions despite years of intensive studies which include using advanced techniques, such as dynamical cluster approximation and quantum Monte Carlo simulation[15](as a Cluster solver),etc. Moreover, there are a number of issues related to cuprates which do not completely fit within the single-band Hubbard model framework. For example, the cuprate gap is set by the charge transfer energy separating the copper and oxygen orbitals[16,17] as opposed to a Mott gap between copper d-states split by the on-site repulsion U. The metal-insulator transition (MIT) seems [18] to be difficult to explain in this framework at any finite U in two-dimensions and higher. These are the reasons for considering a three-band extended Hubbard model(or d-p model[19,20]). The p-and d-fermion nearest neighbor (NN) hoppings $t_{pp}$ and $t_{dd}$ and inter-site, partially screened coulomb interaction $U_1$ for d-fermions are introduced additionally as all the underlying d-p parameters are known[21] to play significant role. The interactions $U_p$ and $U_{dp}$ will not be considered here. In an investigation to search for a hierarchy of multiple many- body interaction scales in high-$T_c$ superconductor, as suggested by recent experimental findings[22-25], these interactions will be taken into account. In momentum space, $U_1$, assumed to be effective for nearest-neighbor(NN) only, corresponds to, say, $U_1(k,q)$ for transition from a momentum q to k.The hopping terms generate k-dependence of the onsite en-ergies. The interaction $U_1(k,q)$ is taken to be separable and expanded in terms of basis functions corresponding to the mixed symmetry states involving $d_{xy}$ and $d_{x^2-y^2}$. The main aim here is to show that in d-p model the non-fermi liquid behavior is generally prevalent once the pure $d_{x^2-y^2}$ wave singlet superconducting instability sets in. A preliminary study, involving the charge and spin ordering gaps in the normal state single-particle spectrum, taking $d_{xy}$ symmetry state into account is the other aim. The motivation is to show in future that anomalous pairing for $(d_{x^2-y^2}+d_{xy})$ wave symmetry possibly holds the explanation regarding the difference in the properties of the nodal and anti-nodal quasi-particle states alluded to above.

The proximity of a quantum critical point (QCP) for a metallic system generates a non Fermi-liquid (NFL) characterized by anomalous (infinite) temperature dependence. Such a NFL behavior has been identified in some cuprates (YBCO, BISCO,...) and in some heavy-fermion materials like ( Ce, Au)$Cu_6$. In particular, the existence of a QCP can explain the crossover from a NFL behavior to a typical Fermi-liquid (FL) behavior. This crossover is observed at optimal doping in high-$T_c$ materials. One of the future aims[48,49] of the investigation is to deal with the FL-NFL crossover aspect in d-p model. Now additional holes are expected to occupy oxygen sites for the hole-doped ($\delta>0$) cuprates. This implies that the renormalized charge transfer gap $\Delta_r$ tends towards zero for $\delta >0$. It will be shown here that the non-Fermi liquid behavior emerges ( for $\Delta_r \rightarrow 0$) once the pure $d_{x^2-y^2}$ wave singlet superconducting instability sets in. The consequence of the fact is that the system is unstable towards a non-BCS pairing. The lifetime ratio of nodal and anti-nodal quasi-particles will also be calculated here. The investigation yields the strong evidence that the latter ones are incoherent. The charge and the spin ordering gaps appear in the single-particle excitation spectrum when $d_{xy}$ component is taken into account. In the mixed-symmetry state, $d_{x^2-y^2}$ and $d_{xy}$ gaps coexist. The investigation also yields two distinct gaps in the normal state spectrum; the larger one corresponds to the sum and the smaller one to the difference of spin and charge ordering gaps. This is expected to shed light on the pseudo-gap phenomenon in the normal state and multi-gap scenario of the cuprates.

The work is organized as follows: In Sec.II the momentum space, three-band Hubbard Hamiltonian (or d-p Hamiltonian) in the mean field approximation involving nearest neighbor(NN) hoppings and the interaction $U_1(\mathbf{k'},\mathbf{k})$(for transition from a momentum $\mathbf{k'}$ to $\mathbf{k}$) corresponding to unscreened, inter-site coulomb repulsion will be presented using a variant of the slave-boson formalism of Kotliar and Ruchenstein[26](KR). In section III it will be shown that the system generally exhibits non-Fermi liquid (NFL) character involving non-BCS gap equation once the $d_{x^2-y^2}$ wave singlet superconducting instability sets in. The lifetime ratio of nodal and anti-nodal quasi-particles will also be calculated in section III to show that the latter ones are incoherent. In sectionIV we show that the charge and the spin ordering gaps appear in the single-particle spectrum when $d_{xy}$ component is taken into account. The work ends with concluding remarks in section V.

**II. THREE-BAND EXTENDED HUBBARD MODEL**

The particle hopping is usually accompanied by non-local disturbances in an interacting system which show up as the band alteration factor in the single-particle spectrum. In the KR approach [26] rewriting the Hubbard Hamiltonian, in terms of original fermion fields and a set of four bosons corresponding to each of the four states available for hopping, is the essential step in this direction. Subsequently, Balseiro et al [19] applied the same formalism for the three-band extended Hubbard model (d-p model[19,20]). The model Hamiltonian with p-and d-fermion nearest neighbor (NN) hoppings $t_{pp}$ and $t_{dd}$ and inter-site, partially screened coulomb interaction $U_1(i,j)$ for d-fermions is given by

$$H = \sum_{i\sigma} \varepsilon_d \, d^{\dagger}_{i\sigma} d_{i\sigma} + \sum_{\langle ij \rangle \sigma} (t_{dd} \, d^{\dagger}_{i\sigma} d_{j\sigma} + h.c.) + \sum_{i\alpha\sigma} \varepsilon_p \, p^{\dagger}_{i+\alpha(a/2),\sigma} p_{i+\alpha(a/2),\sigma}$$

$$+ \sum_{i\alpha\alpha'\sigma;\alpha\neq\alpha'} t_{pp}[- p^{\dagger}_{i+\alpha(a/2)\sigma} p_{i+\alpha'(a/2),\sigma} - p^{\dagger}_{i-\alpha(a/2)\sigma} p_{i-\alpha'(a/2),\sigma} + p^{\dagger}_{i+\alpha(a/2)\sigma} p_{i-\alpha'(a/2),\sigma}$$

$$+ p^{\dagger}_{i-\alpha(a/2)\sigma} p_{i+\alpha'(a/2),\sigma} \,]$$

$$+ \sum_{i\alpha\sigma} t_{pd}[\{ d^{\dagger}_{i\sigma} (p_{i+\alpha(a/2),\sigma} - p_{i-\alpha(a/2),\sigma}) \} + h.c.] + U_d \sum_i n_{i\uparrow} n_{i\downarrow} + \sum_{ij\sigma\sigma'} U_1 n_{i\sigma} n_{j\sigma'}. \quad (1)$$

For $d_{x^2-y^2}$ wave, the interaction $U_1$ will have a form-factor

$$f_d(r) = [\partial (r_y,0) \{ \partial (r_x,a) + \partial (r_x,-a) \} - \partial (r_x,0) \{ \partial (r_y,a) + \partial (r_y,-a) \}] \quad (2)$$

with $\partial (r_{x(y)},0(\pm a))$ being Kronecker's delta. Here spin degeneracy $N \geq 2$, $\alpha \rightarrow (x,y)$, and 'a' is the lattice constant; $n_{i\sigma} = d^{\dagger}_{i\sigma} d_{i\sigma}$ is the occupation number of hole with spin σ at Cu $3d_{x2-y2}$ orbital with i labeling Cu-site. $p^{\dagger}_{i+\alpha(a/2),\sigma}$ create a hole with spin σ on O 2p site $(i+\alpha(a/2))$ while $(\varepsilon_d, \varepsilon_p)$ are constant on-site energies; $t_{pd}$ hybridizes nearest neighbor Cu $3d_{x^2-y^2}$ and $O_{2p}$ orbitals. We have a tight binding picture here – a $p_x$- and a $p_y$- orbital on each oxygen couple with a single $d_{x^2-y^2}$ on the Cu. $U_d$ term gives an energy cost to double occupancy by holes. In this model the $p_y$- orbital for the oxygen lying on a horizontal row of the $CuO_2$ plane does not bond with Cu and neither does the $p_z$ orbital (for an oxygen) in a vertical row.

A variant of the slave boson mean field theoretic approach[26] proposed by Kotliar and Ruchenstein (KR) to study the metal-insulator transition (MIT) was used earlier[47] introducing only two bose fields φ and ψ corresponding to empty and doubly occupied sites respectively and make the replacement of Cu-site operators above by $d^{\dagger}_{i\sigma} \to \acute{a}^{\dagger}_{i\sigma}$ =$(d^{\dagger}_{i\sigma} \varphi_i + sgn(\sigma) \psi^{\dagger}_i d_{i,-\sigma})$. It was shown that $\{\acute{a}_{i\sigma}, \acute{a}^{\dagger}_{i\sigma}\} = 1$ provided $\hat{O}_i = \varphi^{\dagger}_i \varphi_i + \psi^{\dagger}_i \psi_i + \sum_{\sigma} d^{\dagger}_{i\sigma} d_{i\sigma} = 1$. In view of the results above one can write $H'(=(H - \mu \tilde{n})$, where μ is the chemical potential for fermion number) as

$$H' = \sum_{i\sigma} \varepsilon^0_d d^{\dagger}_{i\sigma} d_{i\sigma} + \sum_{i\alpha\sigma} \varepsilon^0_p p^{\dagger}_{i+\alpha(a/2),\sigma} p_{i+\alpha(a/2),\sigma} + (U_d - 2\mu)\sum_i \psi^{\dagger}_i \psi_i$$
$$+ \sum_I \lambda_i (\hat{O}_i - q_0 N) + (t/\sqrt{N}) \sum_{i \eta \sigma} sgn(\eta) (\acute{a}^{\dagger}_{i\sigma} p_{i(\eta),\sigma} + p^{\dagger}_{i(\eta),\sigma} \acute{a}_{i\sigma}) + H'_c. \quad (3)$$

$$\tilde{n} = \sum_{i\sigma} d^{\dagger}_{i\sigma} d_{i\sigma} + \sum_{i\alpha\sigma} p^{\dagger}_{i+\alpha(a/2),\sigma} p_{i+\alpha(a/2),\sigma} + 2\sum_i \psi^{\dagger}_i \psi_i. \quad (4)$$

Here $\varepsilon^0_d = \varepsilon_d - \mu$, $\varepsilon^0_p = \varepsilon_p - \mu$, $i(\eta) = i + \eta(a/2)$, and $\eta = (\pm x, \pm y)$. The term $H'_c$ involves hopping terms and the inter-site coulomb interaction $U_1$. The index σ runs from 1 to N. Following Millis and Lee[27] we have made the replacement $t_{pd} \to t/\sqrt{N}$. The system of interest is, however, $q_0 = ½$ and N=2. In the physical subspace defined by $n_{i\uparrow} n_{i\downarrow} = \psi^{\dagger}_i \psi_i$ and the constraint $\hat{O}_i = 1$, H' has the same matrix elements as those calculated for H in the original Hilbert space; $\lambda_i$ is site-dependent Lagrange multiplier enforcing the constraint. In the mean-field approximation, one replaces the bosonic operators φ and ψ by c-numbers, says, $e_0$ and $D_0$ respectively which have to be self-consistently determined. Moreover, Lagrange multipliers are taken to be same for all sites. One obtained the mean field equations to determine $(e_0, D_0, \lambda)$ in ref.47 along these lines. As regards the equation to determine the chemical potential μ, it may be noted that in the FL state of normal metals the Luttinger theorem[28] (LT) is satisfied. However, since superconductivity is to be investigated here and $\mu = \mu(\Delta_{sc})$(where $\Delta_{sc}$ is the SC gap) there is slight discrepancy[29] between the true fermion number and the volume enclosed by the Fermi surface. One thus obtains an approximate equation for μ here applying LT. We now set the stage to obtain single-particle excitation spectrum writing the momentum space counterpart of (3) in mean field approximation (MFA).

The situation when the spin degeneracy is reduced to N=2 is worth investigating for obvious reason. In momentum space, for δ (δ > 0) away from half-filling, the MF d-p Hamiltonian involving only bose mean-field values $(e_0, D_0)$ (of $\varphi$ and $\psi$), NN hoppings $t_{pp}$ and $t_{dd}$, and the interaction $U_1(\mathbf{k'},\mathbf{k})$ can be expressed as follows:

$$H_{m,d-p} = C + H_{MF}^{(0)} + H_{MF}^{(I)} \tag{5}$$

$$C = N_s N \lambda (e_0^2 + D_0^2 - q_0) + N_s N(U_d - 2\mu)D^2 + ((\varepsilon - \mu)NN_s/2)(1 + \delta - 4D_0^2), \tag{6}$$

$$H_{MF}^{(0)} = \sum_{k\sigma} \varepsilon_d(k) d^\dagger_{k\sigma} d_{k\sigma} + \sum_{k\alpha\alpha'\sigma; \alpha\neq\alpha'} \varepsilon_p(k) p^\dagger_{k\alpha\sigma} p_{k\alpha'\sigma}$$
$$+ \sum_{k\eta\sigma}(2i\, t\, \sin(k_\eta a/2)) \{\acute{\alpha}^\dagger_{k\sigma} p_{k\eta\sigma} - p^\dagger_{k\eta\sigma} \acute{\alpha}_{k\sigma}\} \tag{7}$$

$$H_{MF}^{(I)} = \sum_{k\sigma} \Delta_s(k) d^\dagger_{k+Q,-\sigma} d_{k\sigma} + \sum_{k\sigma} \Delta^\dagger_s(k) d^\dagger_{k\sigma} d_{k+Q,-\sigma} + \sum_{k\sigma} \Delta_c(k) d^\dagger_{k+Q,\sigma} d_{k\sigma}$$
$$+ \sum_{k\sigma} \Delta^\dagger_c(k) d^\dagger_{k,\sigma} d_{k+Q,\sigma} + \sum_{k\sigma} \Delta_{sc}(k) d^\dagger_{k,\sigma} d^\dagger_{-k+Q,-\sigma} + \sum_{k\sigma} \Delta^\dagger_{sc}(k) d_{-k+Q,-\sigma} d_{k,\sigma} \tag{8}$$

$$\acute{\alpha}_{k\sigma} = (e_0 d_{k\sigma} + D_0 \text{sgn}(\sigma) d^\dagger_{-k,-\sigma}), \quad \acute{\alpha}^\dagger_{k\sigma} = (e_0 d^\dagger_{k\sigma} + D_0 \text{sgn}(\sigma) d_{-k,-\sigma}) \tag{9}$$

$$\varepsilon_d(k) = -2t_{dd}(\cos k_x a + \cos k_y a), \quad \varepsilon_p(k) = -4t_{pp} \sin(k_x a/2) \sin(k_y a/2) \tag{10}$$

$$\Delta_s(k) = -\sum_{k'} U_1(\mathbf{k'},\mathbf{k}) \langle d^\dagger_{k',\sigma} d_{k'+Q,-\sigma}\rangle, \quad \Delta^\dagger_s(k) = -\sum_{k'} U_1(\mathbf{k'},\mathbf{k}) \langle d^\dagger_{k'+Q,-\sigma} d_{k',\sigma}\rangle \tag{11}$$

$$\Delta_c(k) = -\sum_{k'} U_1(\mathbf{k'},\mathbf{k}) \langle d^\dagger_{k',\sigma} d_{k'+Q,\sigma}\rangle, \quad \Delta^\dagger_c(k) = -\sum_{k'} U_1(\mathbf{k'},\mathbf{k}) \langle d^\dagger_{k'+Q,\sigma} d_{k',\sigma}\rangle \tag{12}$$

$$\Delta_{sc}(k) = \sum_{k'} U_1(\mathbf{k'},\mathbf{k}) \langle d_{-k'+Q,-\sigma} d_{k',\sigma}\rangle, \quad \Delta^\dagger_{sc}(k) = \sum_{k'} U_1(\mathbf{k'},\mathbf{k}) \langle d^\dagger_{k',\sigma} d^\dagger_{-k'+Q,-\sigma}\rangle \tag{13}$$

Here $\varepsilon_d(\mathbf{k})$ satisfies the perfect nesting condition $\varepsilon_d(\mathbf{k+Q}) = -\varepsilon_d(\mathbf{k})$ with $\mathbf{Q} = (\pm\pi, \pm\pi)$. The d-wave symmetry implies that $U_1(\mathbf{k'},\mathbf{k}) = -U_1(\mathbf{k'},\mathbf{k+Q})$ or $\Delta_{sc}(\mathbf{k+Q}) = -\Delta_{sc}(\mathbf{k})$.

For hole doped (δ >0) cuprates, the additional holes are expected to occupy oxygen sites. One may assume then $\varepsilon_d + \lambda \approx \varepsilon_p \approx \varepsilon$, i.e. the renormalized charge-transfer gap $\Delta_r \to 0$. Furthermore, since for δ away from half-filling the number of available carriers is $(NN_s/2)(1+\delta)$ (where $N_s$ is the number of unit cells), it follows from (4) that the on-site energy terms may be approximated by the c-number $(\varepsilon NN_s/2)(1+\delta-4D_0^2)$ and relegated to Eq.(6). In the next two sections NFL behavior, spin and charge orderings will be examined using this approximate mean-field Hamiltonian $H_{m,d-p}$. The thermal averages involved in $H_{m,d-p}$ can be determined in a self-consistent manner.

## III. NODAL AND ANTINODAL QUASIPARTICLE STATES

At microscopic level, to capture the essential physics of high-$T_c$ cuprates, the Hubbard model, Hubbard-Holstein model, etc. have been examined by previous workers[30-34] with moderate success for variety of reasons. To name a few, violation of Luttinger theorem, break down of fluctuation exchange approximation in the strong coupling regime, order parameter undergoing a transition to s-wave superconductivity (SC) from d-SC away from half-filling in a quasi 2D Holstein model, and so on. Moreover, the widely accepted microscopic models for superconductivity (SC), namely BCS theory and more advanced Migdal-Eliashberg formalism, are implicitly based on the assumptions of a FL normal state. In this section it will be shown how $d_{x^2-y^2}$-wave SC instability involving non-BCS gap equation can develop in the d-p model at low temperature yielding the possibility that antinodal excitations in the underdoped regime are incoherent- a signature of NFL behavior. In the next section we propose a viable scenario for the NFL in the normal state. Though several attempts [35-38] have been made in the past to explain the NFL behavior in the normal state through different routes, such as the marginal Fermi liquid approach of Varma and coworkers[35] and the dual-theoretic approach of Tesanovic[38] and so on, none had been entirely satisfactory. For example in Tesanovic's formulation, the FL picture is actually outside the scope of the theory and a finite doping in this dual theory is likely to appear as a finite magnetic field.

The unscreened, inter-site coulomb interaction in the present tight-binding model is taken to be separable, and expanded in terms of some general basis functions $\acute{\eta}_{ik}$, labeled by index i, so that $U_1(k',k) = \sum_i g_i \, \acute{\eta}_{ik'} \acute{\eta}_{ik}$. Here $g_i$ is the coupling of the effective interaction in the specific angular momentum state specified by the index i above. In two dimensions we have $\acute{\eta}_{ik} = \acute{\eta}_{ik}(k_x, k_y)$, where for $d_{x^2-y^2}$ wave $\acute{\eta}_{1k} = (\cos k_x a - \cos k_y a)$, for $d_{xy}$ wave $\acute{\eta}_{2k} = (2 \sin k_x a \cdot \sin k_y a)$ and so on. The orthogonality property of functions $\acute{\eta}_{ik}$ is given by $\sum_k \acute{\eta}_{ik} \acute{\eta}_{jk} = 0$ for $i \neq j$. Now although $d_{x^2-y^2}$ wave pairing is universally recognized[1-4] as the dominant mechanism of high temperature superconductivity, an additional order parameter of $d_{xy}$–wave type is possibly required to explain the relevant experimental findings[39-43]. Moreover, although the experimental evidences for a NFL normal state(NS) in the cuprates are so compelling and universally recognized, a satisfactory microscopic foundation of such a state is not yet available. For these reasons, we investigate here symmetry states involving $d_{x^2-y^2}$ and $d_{xy}$. To this end, we write $U_1(\mathbf{k'},\mathbf{k}) = (g_1 \acute{\eta}_{1k'} \acute{\eta}_{1k} + g_2 \, \acute{\eta}_{2k'} \acute{\eta}_{2k})$ with the aim to give a consistent microscopic foundation of NFL NS.

We consider pure $d_{x^2-y^2}$ wave state first. Since the Hamiltonian $H_{m,d-p}$ is completely diagonal one can write down the equations for the operators $\{d_{k\sigma}(\tau), d^{\dagger}_{-k+Q,-\sigma}(\tau), p_{k\alpha\sigma}(\tau), p^{\dagger}_{-k,\eta,-\sigma}(\tau), \ldots \}$, where the time evolution an operator O is given by

$$O(\tau) = \exp(H_{m,d-p}\tau) \; O \; \exp(-H_{m,d-p}\tau), \qquad (14)$$

to ensure that the thermal averages in $H_{m,d-p}$ are determined in a self-consistent manner, as

$$(\partial/\partial\tau) \, d_{k\sigma}(\tau) = -\varepsilon_d(k) \, d_{k\sigma}(\tau) - \sum_\eta (2i \, e_0 t \sin(k_\eta a/2)) \, p_{k\eta\sigma}(\tau)$$
$$+ \sum_\eta (2i \, t \, D_0 \mathrm{sgn}(-\sigma)) \sin(k_\eta a/2)) \, p^\dagger_{-k,\eta,-\sigma}(\tau) - 2\Delta_{sc}(k) \, d^\dagger_{-k+Q,-\sigma}(\tau), \qquad (15)$$

$$(\partial/\partial\tau) \, d^\dagger_{-k+Q,-\sigma}(\tau) = -\varepsilon_d(k)) \, d^\dagger_{-k+Q,-\sigma}(\tau) - \sum_\eta (2i \, e_0 t \cos(k_\eta a/2)) \, p^\dagger_{-k+Q,\eta,-\sigma}(\tau)$$
$$+ \sum_\eta (2i \, t \, D_0 \mathrm{sgn}(\sigma) \cos(k_\eta a/2)) \, p_{k,\eta,\sigma}(\tau) - 2\Delta^\dagger_{sc}(k) \, d_{k\sigma}(\tau), \qquad (16)$$

$$(\partial/\partial\tau) \, p_{k\eta\sigma}(\tau) = -\varepsilon_p(k) \, p_{k\alpha\sigma}(\tau) + (2i \, t \sin(k_\eta a/2)) \, \acute{a}_{k\sigma}(\tau), \qquad (17)$$

$$(\partial/\partial\tau) \, p^\dagger_{-k,\eta,-\sigma}(\tau) = -\varepsilon_p(k) \, p^\dagger_{-k,\eta,-\sigma}(\tau) - (2i \, t \sin(k_\eta a/2)) \, \acute{a}^\dagger_{k\sigma}(\tau). \qquad (18)$$

For pure $d_{x^2-y^2}$ wave symmetry, the charge and spin ordering gaps (see Eqs.(11) and (12)) are conspicuous by their absence in Eqs.(15) - (18) for the reason that $\Delta(\mathbf{k+Q}) = -\Delta(\mathbf{k})$. The Green's function $G_{dd}(k,\sigma,\tau) = -\langle T\{d_{k\sigma}(\tau) d^\dagger_{k\sigma}(0)\}\rangle$, where T is the time-ordering operator which arranges other operators from right to left in the ascending order of time $\tau$, is of primary interest as the poles of the Fourier transform of this function yields the single-particle excitation spectrum. The other thermal averages of interest are

$$G^{(a)}_{dd}(-k+Q,-\sigma,\tau) = -\langle T\{d^\dagger_{-k+Q,-\sigma}(\tau) \, d^\dagger_{k\sigma}(0)\}\rangle, \qquad (19)$$

$$G^{(a)}_{pd,\eta}(-k,\eta,-\sigma,\tau) = -\langle T\{p^\dagger_{-k,\eta,-\sigma}(\tau) \, d^\dagger_{k\sigma}(0)\}\rangle, \qquad (20)$$

$$G_{pd,\eta}(k,\eta,\sigma,\tau) = -\langle T\{p_{k,\eta,\sigma}(\tau) \, d^\dagger_{k\sigma}(0)\}\rangle, \qquad (21)$$

as these are required for setting up a system of equations to determine the anomalous pairing gap $\Delta_{sc}(\mathbf{k}) = \sum_{\mathbf{k'}} U_1(\mathbf{k'},\mathbf{k}) \langle d_{-\mathbf{k'}+Q,-\sigma} d_{\mathbf{k'},\sigma}\rangle$ in the excitation spectrum. For a pure $d_{x^2-y^2}$ gap function, in view of $U_1(\mathbf{k'},\mathbf{k}) = (g_1 \acute{\eta}_{1k'} \acute{\eta}_{1k} + g_2 \acute{\eta}_{2k'} \acute{\eta}_{2k})$, one may write $\Delta_{sc}(\mathbf{k}) = \Delta_0 (\cos k_x a - \cos k_y a)$; $\Delta_{sc}(\mathbf{k})$ vanishes linearly in the four nodes, i.e. $\mathbf{k}$ parallel to $(\pm\pi, \pm\pi)$. The Fourier coefficients of the thermal averages above are the Matsubara propagators $\{G_{dd}(k, \sigma, z), G^{(a)}_{dd}(-k+Q, -\sigma, z), G^{(a)}_{pd,\eta}(-k, \eta, -\sigma, z), G_{pd,\eta}(k, \eta, \sigma, z)\}$ where $z = [(2n+1)\pi i / \beta]$ with $n = 0, \pm 1, \pm 2, \ldots$. With the aid of Eqs.(15)-(18) one obtains the following equations of these propagators:

$$(z - \varepsilon_d(k)) G_{dd}(k, \sigma, z) + (-\sum_\eta (2i e_0 t \sin(k_\eta a/2)) G_{pd,\eta}(k, \eta, \sigma, z) +$$
$$(\sum_\eta (2i t D_0 \mathrm{sgn}(-\sigma) \sin(k_\eta a/2))) G^{(a)}_{pd,\eta}(-k, \eta, -\sigma, z) - 2\Delta_{sc}(k) G^{(a)}_{dd}(-k+Q, -\sigma, z) = 1, \quad (22)$$

$$(-2i t D_0 \mathrm{sgn}(-\sigma) \sin(k_\eta a/2)) G_{dd}(k, \sigma, z) + (z - \varepsilon_p(k)) G^{(a)}_{pd,\eta}(-k, \eta, -\sigma, z) = 0, \quad (23)$$

$$(2i e_0 t \sin(k_\eta a/2)) G_{dd}(k, \sigma, z) + (z - \varepsilon_p(k)) G_{pd,\eta}(k, \eta, \sigma, z) = 0, \quad (24)$$

$$-2\Delta^\dagger_{sc}(k) G_{dd}(k, \sigma, z) + \sum_\eta (2i t D_0 \mathrm{sgn}(\sigma) \cos(k_\eta a/2)) G_{pd,\eta}(k, \eta, \sigma, z) +$$
$$(-\sum_\eta (2i e_0 t \cos(k_\eta a/2)) G^{(a)}_{pd,\eta}(-k, \eta, -\sigma, z) + (z - \varepsilon_d(k)) G^{(a)}_{dd}(-k+Q, -\sigma, z) = 0. \quad (25)$$

It is tedious but straightforward to see from (23)-(25) that the Fourier coefficient $G_{dd}(k, \sigma, z) = (X^2 Y / XD)$, where $X = (z - \varepsilon_p(k))$, $Y = (z - \varepsilon_d(k))$, $D = (XY^2 - 4Y\zeta^2 - 4X\Delta^2_{sc}(k))$, and $\zeta^2 = t^2(e_0^2 + D_0^2) \sum_{k\eta} \sin^2(k_\eta a/2)$. Near metal-insulator transition (MIT), the quantities $(e_0^2, D_0^2)$ tends towards zero[26]. Therefore, away from MIT in the limit $(k_B T)$ larger compared to the hopping terms, one finds $D \approx Y(XY - 4(\zeta^2 + \Delta^2_{sc}(k)))$. In this assumed situation one obtains an FL scenario characterized by single-particle Green's function endowed with simple poles at $(\varepsilon_1(k), \varepsilon_2(k), \varepsilon_p(k), \varepsilon_d(k))$ where

$$\varepsilon_1(k) = (1/2)(\varepsilon_p(k) + \varepsilon_d(k) - E_k), \quad \varepsilon_2(k) = (1/2)(\varepsilon_p(k) + \varepsilon_d(k) + E_k), \quad (26)$$

$$E_k = \sqrt{\{G^2 + 16(\zeta^2 + \Delta^2_{sc}(k))\}}, \quad G = (\varepsilon_p(k) - \varepsilon_d(k)). \tag{27}$$

The location of these poles in the complex plane defines unambiguously the energy-momentum relation of the quasi-particle eigenstate and their lifetime, assumed to be long enough. Away from MIT in the small temperature limit, the NFL behavior is prevalent. There are four poles of the function $G_{dd}(k, \sigma, z)$ at frequencies $z = \{\varepsilon_p(k), \varepsilon_i(k)\}$ with $i = 1, 2, 3$; $\varepsilon_i(k)$ being given by the roots of the equation $D = 0$. This equation may be written in a slightly different form $z'^3 - 3a\,z' - 2h_1 = 0$, where $z' = z + (q/(3p))$, $p = 1$, $q = -(\varepsilon_p(k) + 2\varepsilon_d(k))$, $r = [(2\varepsilon_d(k)\varepsilon_p(k) + \varepsilon^2_d(k)) - 4(\zeta^2 + \Delta^2_{sc}(k))]$, $s = [4\{\varepsilon_d(k)\zeta^2 + \varepsilon_p(k)\Delta^2_{sc}(k)\} - \varepsilon^2_d(k)\varepsilon_p(k)]$, $a = [(q^2/9p) - (r/3)]$, and $h_1 = \{-(q^3/(27p^2)) + ((qr)/(6p)) - (s/2)\}$. The cubic in $z'$, for $a^3 > h_1^2$, has three admissible solutions

$$\text{Re } z' = 2a^{1/2} \cos(\tilde{a}/3), \quad -2a^{1/2} \cos((\pi+\tilde{a})/3), \quad -2a^{1/2} \cos((\pi-\tilde{a})/3) \tag{28}$$

and Im $z' = \pm(3a)^{1/2} \cos^{1/2}(2\tilde{a}/3)$, where $\cos \tilde{a} = -h_1/(a^3)^{1/2}$. For many years, it has been recognized that the properties of the quasi-particle states, such as the quasi-particle life-time (QPLT) etc., near the nodes of the d-wave superconducting gap in the cuprates are quite different from those near the gap maxima or antinodes. It will now be shown that the nodal QPLTs are significantly longer than those for antinodal quasiparticles. The lifetime ($\tau$) of quasiparticles is given by $(1/\tau) = |\text{Im } z|$. In view of $\Delta_{sc}(\mathbf{k}) = \Delta_0 (\cos k_x a - \cos k_y a)$ and the results above one finds that

$$1/\tau_{\text{antinode}} = 2(\zeta'^2 + 4\Delta^2_0)^{1/2}, \quad 1/\tau_{\text{node}} = 2(\zeta'^2 + (t^2_{pp}/3))^{1/2} \cos^{1/2}(2\tilde{a}_1/3). \tag{29}$$

Here $\zeta'^2 = t^2(e_0^2 + D_0^2)$ is the dressed hybridization parameter, and

$$\cos \tilde{a}_1 = 3\sqrt{3}\{(4/3)t_{pp}\zeta'^2 + (8/27)t^3_{pp}\}/\sqrt{\{(4t^2_{pp} + 12\zeta'^2)^3\}}. \tag{30}$$

Away from MIT deep in the superconducting (SC)phase, one finds in view of Eqs.(29) and (30) that for $(t^2_{pp}/\zeta'^2) = (0.1000$ to $0.3000)$ and $(t^2_{pp}/\Delta^2_0) = 0.1000$ and $0.0625$, $(\tau_{node}/\tau_{antinode}) \sim 5$. This implies that the nodal quasi-particles are sharper and have longer lifetime than the anti-nodal ones. This dichotomy, where "hot"(antinodal) quasi-particle becomes insulating while "cold" (nodal) quasi-particles remain metallic, exists only when one stays away from SC transition region. In the proximity, this feature disappears. In fact, the anti-nodal quasi-particles in this case have longer lifetime than the nodal ones ( which can be verified from (29)taking ,say, $(t^2_{pp}/\zeta'^2) = 1$ to $100$ and $(t^2_{pp}/\Delta^2_0) = 4- 40)$.One explanation, for the "dichotomy", that has been proposed is the existence of intense magnetic scattering near the antinodal point in the underdoped materials due to the proximity of the wave vector Q connecting the anti-nodal points by an anti-ferromagnetic nesting vector. This scattering has also recently been related by several authors to the properties of the Fermi surface in the pseudo-gap state[44-46]. We find that the reason for the dichotomy is the non-validity of FL picture which , in turn, is due to the quantum critical point (QCP) proximity discussed in ref.48.

## IV. SPIN AND CHARGE ORDERINGS

For many years, it has been recognized that the properties of the quasiparticle states near the nodes of the superconducting gap in the cuprates are quite different from those near the gap maxima or antinodes. In fact, both transport and ARPES nodal mean free paths are significantly longer than those extracted from ARPES for anti-nodal quasi-particles. In the last section we have shown that the anomalous pairing of d-fermion fields is possible for pure $d_{x^2-y^2}$ wave symmetry leading to a d-wave gap.

The charge and spin ordering gaps are conspicuous by their absence in the single-particle spectrum in this case. In this section, these gaps will be shown to appear when $d_{xy}$ component is taken into account. The motivation behind the exercise, as already stated, is to set the stage for investigating in future the anomalous pairing involving $(d_{x^2-y^2}+d_{xy})$ wave symmetry, for this type of pairing seemingly holds the explanation regarding the difference in the properties of the nodal and anti-nodal quasi-particle states.

Since the Hamiltonian $H_{m,d-p}$ is completely diagonal one can write down the equations of motion of the operators { $d_{k\sigma}(\tau)$, $d_{k+Q,-\sigma}(\tau)$, $d_{k+Q,\sigma}(\tau)$,…. } easily, where the time evolution of an operator is given by Eq.(17). The thermal averages of interest are

$$
\begin{aligned}
G_{dd}(k, \sigma, \tau) &= -\langle T\{ d_{k\sigma}(\tau)\, d^{\dagger}_{k\sigma}(0)\}\rangle, \\
G^{(s)}_{dd}(k+Q,-\sigma, \tau) &= -\langle T\{ d_{k+Q,-\sigma}(\tau)\, d^{\dagger}_{k\sigma}(0)\}\rangle, \\
G^{(c)}_{dd}(k+Q, \sigma, \tau) &= -\langle T\{ d_{k+Q,\sigma}(\tau)\, d^{\dagger}_{k\sigma}(0)\}\rangle, \\
G^{(a)}_{pd,\eta}(-k,\eta,-\sigma, \tau) &= -\langle T\{ p^{\dagger}_{-k,\eta,-\sigma}(\tau)\, d^{\dagger}_{k\sigma}(0)\}\rangle, \\
G_{pd,\eta}(k, \eta, \sigma, \tau) &= -\langle T\{ p_{k,\eta,\sigma}(\tau)\, d^{\dagger}_{k\sigma}(0)\}\rangle
\end{aligned}
\tag{31}
$$

as these are required for setting up a system of equations to determine the gaps $\Delta_s(\mathbf{k}) = -\sum_{\mathbf{k'}} U_1(\mathbf{k'},\mathbf{k}) \langle d^{\dagger}_{\mathbf{k'},\sigma} d_{\mathbf{k'+Q},-\sigma}\rangle$ and $\Delta_c(\mathbf{k}) = -\sum_{\mathbf{k'}} U_1(\mathbf{k'},\mathbf{k}) \langle d^{\dagger}_{\mathbf{k'},\sigma} d_{\mathbf{k'+Q},\sigma}\rangle$ in the excitation spectrum. The Fourier coefficients of the thermal averages above are the Matsubara propagators { $G_{dd}(k,\sigma,z)$, $G^{(s)}_{dd}(k+Q,-\sigma,z)$, $G^{(c)}_{dd}(k+Q,\sigma,z)$, $G^{(a)}_{pd,\eta}(-k,\eta,-\sigma,z)$, and $G_{pd,\eta}(k, \eta, \sigma, z)$} where $z = [(2n+1)\pi i/\beta]$ with $n = 0, \pm 1, \pm 2,\ldots$. As in section III, one obtains the following equations of the coefficients:

$(z - \varepsilon_d(k))\, G_{dd}(k,\sigma, z) + (-\sum_\eta (2i\, e_0\, t\, \sin(k_\eta a/2)))\, G_{pd,\eta}(k, \eta, \sigma, z) +$

$(\sum_\eta (2i\, t\, D_0 sgn(-\sigma) \sin(k_\eta a/2))) G^{(a)}_{pd,\eta}(-k,\eta,-\sigma,z) - 2\Delta_s(k)\, G^{(s)}_{dd}(k+Q,-\sigma,z)$

$$-2\Delta_c(k)\, G^{(c)}_{dd}(k+Q,\sigma,z) = 1, \qquad (32)$$

$(-2i\, t\, D_0 sgn(-\sigma) \sin(k_\eta a/2))\, G^{(a)}_{pd,\eta}(k,\eta,\sigma,z) + (z - \varepsilon_p(k))\, G^{(a)}_{pd,\eta}(-k,\eta,-\sigma,z)=0, \quad (33)$

$(2i\, e_0 t \sin(k_\eta a/2))\, G_{dd}(k,\sigma,z) + (z - \varepsilon_p(k))\, G_{pd,\eta}(k,\eta,\sigma,z) = 0, \qquad (34)$

$-2\Delta_s(k)\, G_{dd}(k,\sigma,z) + \sum_\eta (-2i\, t\, D_0 sgn(-\sigma) \sin(k_\eta a/2))\, G^{(a)}_{pd,\eta}(-k,\eta,-\sigma,z)$

$$+ (z - \varepsilon_d(k+Q))\, G^{(s)}_{dd}(k,\sigma,z) - 2\Delta_c(k)\, G_{dd}(k,-\sigma,z) = 0. \qquad (35)$$

$-2\Delta_c(k)\, G_{dd}(k,\sigma,z) + \sum_\eta (-2i\, t\, D_0 sgn(\sigma) \sin(k_\eta a/2))\, G^{(a)}_{pd,\eta}(-k,\eta,\sigma,z)$

$$+ (z - \varepsilon_d(k+Q))\, G^{(c)}_{dd}(k,\sigma,z) - 2\Delta_s(k)\, G_{dd}(k,-\sigma,z) = 0. \qquad (36)$$

$(-\sum_\eta (2i\, e_0 t \sin(k_\eta a/2))\, G_{pd,\eta}(k,\eta,\sigma,z) + \sum_\eta (2i\, t\, D_0 sgn(\sigma) \sin(k_\eta a/2))$

$G^{(a)}_{pd,\eta}(-k,\eta,\sigma,z) - 2\Delta_c(k)\, G^{(s)}_{dd}(k,\sigma,z) - 2\Delta_s(k)\, G^{(c)}_{dd}(k,\sigma,z)$

$$+ (z - \varepsilon_d(k))\, G_{dd}(k,-\sigma,z) = 0. \qquad (37)$$

It is very tedious but straight-forward to see from (32)-(37) that the Fourier coefficient $G_{dd}(k,\sigma,z) = (\Gamma_1(z)/\Gamma(z))$, where $\Gamma(z) = (z - \varepsilon_p(k))\, \gamma(z)$, and

$\gamma(z) = [(z - \varepsilon_p(k))\, (z^2 - \varepsilon^2_d(k))^2 + (A^2+B^2)\, (z^2 - \varepsilon^2_d(k))\, (z + \varepsilon_d(k))$

$+ 2\, (\Delta_s(k) - \Delta_c(k))\{2A^2\Delta_c(k) + B^2(z + \varepsilon_d(k))\}(z + \varepsilon_d(k))$

$- 8\, (z^2 - \varepsilon^2_d(k))\, (z - \varepsilon_p(k))\, (\Delta^2_s(k) + \Delta^2_c(k))$

$+B^2(z + \varepsilon_d(k) + \Delta_s(k) - \Delta_c(k))\, (\Delta_s(k) + \Delta_c(k))^2 + 16(\Delta^2_s(k) - \Delta^2_c(k))^2(z - \varepsilon_p(k))], (38)$

$$A = \sum_\eta (2i\, e_0 t \sin(k_\eta a/2)), \quad B = \sum_\eta (2i\, t\, D_0 \sin(k_\eta a/2)). \qquad (39)$$

The single-particle spectrum is given by the roots of the quintic $\gamma(z)= 0$ and $\varepsilon_p(k)$. This general quintic can be solved, in principle, in terms of Jacobi theta function expressing the former into Bring quintic form. We refrain from giving details, for one possibly does not gain much insight by undertaking this task.

In the proximity of MIT, however, since $(e^2_0, D^2_0) \to 0^+$ one notices that $\Gamma(z)$ is approximately factorizable: $\Gamma \approx (z - \varepsilon_p(k))^2 (z + \alpha_+)(z - \alpha_+)(z + \alpha_-)(z - \alpha_-)$, where $\alpha_+ = \sqrt{\{\varepsilon^2_d(k) + 4((\Delta_s(k) + \Delta_c(k))^2\}}$ and $\alpha_- = \sqrt{\{\varepsilon^2_d(k) + 4((\Delta_s(k) - \Delta_c(k))^2\}}$. For $(e^2_0, D^2_0) \to 0^+$ we find that $\Gamma_1(z) \approx (z - \varepsilon_p(k))^2 [(z + \varepsilon_d(k))(z^2 - \varepsilon^2_d(k) - 4\Delta^2_s(k)) - 4\Delta^2_c(k)(z - \varepsilon_d(k))]$. One is thus able to observe that the normal phase, in the proximity to MIT, is a near Fermi liquid characterized by two gaps $(\Delta_s(k) + \Delta_c(k))$ and $(\Delta_s(k) - \Delta_c(k))$. The former corresponds to a higher energy scale while the latter to a lower one. The result, at least qualitatively, is in agreement with recent experimental findings[22-25].

One of the central issues in a theory of strongly correlated systems is the existence or not of well-defined quasi-particles. This question is best addressed by studying the spectral weight (SW) $A(k, \omega)$. In a Fermi liquid (FL), the SW is dominated by quasi-particle peaks. We find here $A(k, \varepsilon)$ to be a bunch of the $\delta$-function with Bogoliubov coherence factors $\{u_k^{+2}, v_k^{+2}, u_k^{-2}, v_k^{-2}\}$:

$$A(k, \varepsilon) = u_k^{+2} \delta(\varepsilon + \alpha_+) + v_k^{+2} \delta(\varepsilon - \alpha_+) + u_k^{-2} \delta(\varepsilon + \alpha_-) + v_k^{-2} \delta(\varepsilon - \alpha_-)$$

where $(u_k^{+2} + v_k^{+2} + u_k^{-2} + v_k^{-2}) = 1$. The other three equations to determine $(u_k^{+2}, v_k^{+2}, u_k^{-2}, v_k^{-2})$ are

$$- u_k^{+2} \alpha_+ + v_k^{+2} \alpha_+ - u_k^{-2} \alpha_- + v_k^{-2} \alpha_- = \varepsilon_d$$

$$u_k^{+2} \alpha^2_- + v_k^{+2} \alpha^2_- + u_k^{-2} \alpha^2_+ + v_k^{-2} \alpha^2_+ = \varepsilon^2_d + 4(\Delta^2_s(k) + \Delta^2_c(k))$$

$$- u_k^{+2} \alpha_+ \alpha^2_- + v_k^{+2} \alpha_+ \alpha^2_- - u_k^{-2} \alpha_- \alpha^2_+ + v_k^{-2} \alpha_- \alpha^2_+ = \varepsilon_d \{\varepsilon^2_d + 4(\Delta^2_s(k) - \Delta^2_c(k))\}. \quad (40)$$

In order to obtain a smooth spectrum, the $\delta$-functions above could be replaced with a Lorentzian of half-width $\kappa$, e.g. for $\delta(\varepsilon - \alpha_+)$ one can write

$$(\kappa \pi) \delta( \varepsilon - \alpha_+ ) = [\{ ( \varepsilon / \kappa) - (\alpha_+ / \kappa )\}^2 + 1]^{-1}. \tag{41}$$

With this replacement and calculating the factors ($u_k^{+2}$, $v_k^{+2}$, $u_k^{-2}$, $v_k^{-2}$) we obtain an explicit expression for A(k, ε). A quantitative comparison with available experimental findings[22-25] (where the pseudo-gap is reported to have two distinct energy scales: one with a low-energy comparable to that of the superconductivity gap and another one with larger energy) is a future task. The result obtained here suggests good prospects for our approach to unravel the mystery of the normal state.

## V. CONCLUDING REMARKS

We have reported in ref.48 that the insulator-to-metal transition discussed earlier (see ref.47), signaled by the mean field values of the bose fields $D_0$ and $e_0$, respectively, corresponding to doubly occupied and empty sites acquiring non-zero values, is followed by the onset of a quantum fluctuations driven singlet pairing instability (SPI) as the temperature is lowered in the near zero doping limit. The transition temperature could be driven to zero, tuning the parameter 'u' in ref.47, to have access to quantum criticality provided the spin degeneracy N is strictly much greater than one. The quantum fluctuations then get the full opportunity to dominate over thermal fluctuations. In this communication, however, we have investigated the situation when the spin degeneracy is reduced to N=2. It is shown in section III that the system exhibits non-Fermi liquid (NFL) behavior characterized by non-BCS gap equation once the pure $d_{x^2-y^2}$ wave singlet superconducting instability sets in. Furthermore, we find that the normal phase has coherent quasi-particle scenario (i.e. perfect FL picture) in the momentum space close to metal-insulator transition (MIT); elsewhere the non-Fermi liquid behavior is the prevalent one. All these are not surprising, since for N=2 quantum fluctuations never get the full opportunity to dominate over thermal fluctuations and the character of the transition remains quasi-normal.

Finally, the multi-gap scenario obtained here is in qualitative agreement with the multiple-energy scales experimentally observed[3,22-25] in the hole-doped cuprates, implying the observation to be a generic feature of these systems.